# Do grid codes afford generalization and flexible decision-making?

Linda Q. Yu *,◊,a, Seongmin A. Park *,◊,b, Sarah C. Sweigart b,c, Erie D. Boorman b,c,†, Matthew R. Nassar a,†

a. Carney Institute for Brain Sciences, Brown University, Providence, RI, 02912, USA
b. Center for Mind and Brain, University of California, Davis, Davis, CA, USA
c. Department of Psychology, University of California, Davis, Davis, CA, USA

\* These authors contributed equally to this work
◊ Corresponding
† These authors contributed equally to this work

Behavioral flexibility is learning from previous experiences and planning appropriate actions in a changing or novel environment. Successful behavioral adaptation depends on internal models the brain builds to represent the relational structure of an abstract task. Emerging evidence suggests that the well-known roles of the hippocampus and entorhinal cortex (HC-EC) in integrating spatial relationships into cognitive maps can be extended to map the transition structure between states in non-spatial abstract tasks. However, what the EC grid-codes actually compute to afford generalization remains elusive. We introduce two non-exclusive ideas regarding what grid-codes may represent to afford higher-level cognition. One idea is that grid-codes are eigenvectors of the successor representation (SR) learned online during a task. This view assumes that the grid codes serve as an efficient basis function for learning and representing experienced relationships between entities. Subsequently, the grid codes facilitate generalization in novel contexts such as when the goal changes. The second idea is that the grid-codes reflect the inferred global task structure. This view assumes that the grid-code represents a structural code that is factorized from specific sensory content, enabling structural information to be transferred across tasks. Subsequently, the brain could afford one-shot inferences without requiring experience. The ability to generalize experiences and make appropriate decisions in novel situations is critical for both animals and machines. Here we review proposed computations of the grid-code in the brain, which is potentially critical to behavioral flexibility.

# 1. Introduction

Generalization is an important problem in reinforcement learning. It is not efficient for organisms or agents to discard all of their previous experiences and learn everything from scratch every time they encounter a new environment, or a new object. Instead, one would want to transfer some of what had been learnt previously about another environment or object onto this new instance. However, generalization is also very difficult. In which ways is this new object or environment similar to the previous one, and in which ways is it different? What are the relevant factors – in other words, what is the *structure* – we need to pull out in order to generalize?

Having an accurate internal model of task structures is extremely important to afford behavioral flexibility and generalization. Reinforcement learning has been used to explain how the brain learns the values assigned to states or actions over the task space and how to maximize reward. However, maintaining learned values without having an explicit representation of the task structure will cause animals to fail to achieve flexibility when facing changes in the environment. For example, the animal would need to relearn values when the reward location or the task demand is changed. Representational learning, to the contrary, assumes that the animal uses previous experiences to construct an internal model of the relationships between states in an abstract task. Subsequently, it can leverage the model for generalizability and flexibility [1–5]. Moreover, the internal tracking of regularities between experiences facilitates what to learn and how to learn, supporting future learning and effective exploration [6,7].

The medial temporal lobe, especially the hippocampal-entorhinal cortical (HC-EC) system, has been found to represent spatial relationships between landmarks while an animal explores physical space [8–12]. There are at least two possible ways of how the HC-EC system might encode and represent relational structure to generalize previous experiences (Fig.1A). On one hand, the brain might build a representation of successors based on the probability of future occupancy over states based on previous experiences. This "successor representation" allows one to appropriately transfer new information about a reward to the states most predictive of it. Such a mechanism would allow one to learn a new policy and adapt to the changes in the task demand within the same task space, such as changing goal locations (red arrows in Fig.1A). On the other hand, the brain may infer the latent global structure from the stream of events and construct a graphical representation of relationships between states in the abstract task. This internal representation may provide a cognitive map to guide future goal-directed navigation, and allow one to generalize the policy that was optimized in one task to a completely different task which comprises different sensory stimuli, if relationships between states in two different tasks are represented with the same graph structure (Fig.1A). This allows one to make one-shot inferences of unlearned relationships (Blue arrows in Fig.1A and Fig.1C).

Both views could theoretically support various aspects of recent evidence showing that relational structure between states in an abstract task can be embedded in the HC-EC representation, such as a metric space or cognitive graph encoding relationships in both continuous and discrete dimensions [8,9,13–16] While several proposals have been made about how the brain affords the first type of generalization within learned environments [17,18], more studies are required to understand whether and how the brain generalize previous experiences to find solutions for completely novel problems (the second type of generalization), particularly in a conceptual space.

Grid cells have garnered growing interest as a way brains can represent structure. Grid cells were first discovered in the rodent medial entorhinal cortex [8]. The firing pattern of these cells overlaid the spatial arena in which the rodent was navigating in a fascinating tessellating hexagonal pattern. Particularly, grid-like neural patterns have been discovered to respond not to just spatial navigation, but when thinking about non-spatial relationships [19–23]. These findings have prompted a burgeoning of research into what causes these grid-like responses, what factors change their behavior, and what uses they might have for learning and generalization. For instance, place and grid cells have emerged from different computational models [3,17,24–26], but so far there has been limited scholarship scrutinizing the differences between these perspectives and their attendant implications. In this article, we will review and contrast two of the models that have been proposed, which could explain grid cell firing and what implications they would have for learning: eigenvectors of the successor representation (SR) and an abstracted graphical representation of an environment or task's structural form. These two perspectives agree that grid cells represent the structure of an environment and can facilitate generalization. However, they differ in the proposed mechanisms and therefore, in our view, make differing predictions under certain conditions.

## 2. Grid codes as eigenvectors of successor representation

**SR in learning & modelling**

One prominent theory of grid cells is that they are derived from hippocampal place cell representations, many aspects of which can be described with the successor representation [17,27]. The successor representation (SR) is an efficient basis function for reinforcement learning in which "credit" is distributed according to the discounted future occupancy of each state [27]. A state here can be a particular location in space, or a particular position (e.g., a person in a social network, a frequency in a series of pure tones, etc.) in a cognitive space. In an SR, each state comes to predict the one following it, with a discount rate that makes the strengths of the predictions fall off at some point (Fig. 1B, left). Viewed in this way, a place cell's receptive field encodes which locations predict that state. Given a random uniform policy (in other words, the animal can freely roam around in all directions equally), this place cell's receptive field will be circular, because it will predict that the animal can occupy any state next to the current one with equal probability.

Recent theoretical work has suggested that grid cells could reflect eigenvectors of the successor representation (Fig. 1B, right). Eigenvectors of the SR form tessellated patterns that are sparser than the firing patterns themselves, analogous to its principal components. In open fields (given a uniform policy), this representation distributes itself evenly in a hexagonal pattern according to the mathematics of packing, in similar fashion to the firing fields of grid cells [28]. However, this implies that if there exists a less-than-uniform policy, such that the arena is not regularly shaped (as in a trapezoid, for example), if there are obstructions or bottlenecks, or if a reward means that the animal expects to spend more time at the goal location, the uniform tiling prescribed by the SR will be disrupted accordingly [29–32]. Importantly, the policy-dependent nature of SR means that it can't "see through walls", so to speak – it would represent paths that can actually be travelled, instead of Euclidean, as-the-crow-flies distances.

Models using the SR framework have been shown to recreate real-life activity patterns of place and grid cells well [17]. In addition to replicating the asymmetry of grid fields in trapezoidal compared to square environments, the SR model can replicate the evolution with experience grid cell firing in an environment with multiple compartments [17]. An extension of the SR model that incorporated boundary vector cells (inspired by border cells in entorhinal cortex) is able to mimic even more accurately the firing patterns of place and grid cells, such as their elongation along borders and instant representation of inner walls [33]. Similarly, incorporating the multi-scale nature of grid and place modules (where the receptive fields enlarge along the dorsal-ventral axis) allows for multiple discount factors in the SR model, disambiguating between distance and occupancy of states [34].

**SR eigenvectors facilitating generalization transfer of structure**

The standard SR, by itself, cannot generalize very well to novel contexts. This is because it learns the transitions between specific stimuli or states. Accordingly, place cells remap between contexts, breaking the relationships between stimuli. In contrast, we propose that the eigenvector of the SR, which is the mechanism for grid cells in this perspective, should be able to generalize in some circumstances. The principal components of the SR theoretically can serve as a basis function that preserve and transfer learning in a new state. Supporting this view, grid codes form immediately when animal enters a novel space though it takes time to stabilize, which suggest that there is an immediate transfer of a basis set in a novel space that is then improved through further experience [35,36]. Additionally, when changing environments grid cells reorient in a coordinated fashion, thereby preserving the relationships to each other [36].

Two types of generalization should be possible using SR eigenvectors. First, relationships learnt from one part of a space should be able to be generalized to another part of the same space. That is, if we learn about transitions between one subset of stimuli that can be summarized by a set of eigenvectors, and we also learn that the remaining subset of stimuli experience some of the same transitions, then the SR eigenvectors would be able to "fill out" the rest of the transitions, even though we have never directly experienced them (Fig. 2A). This is because eigenvectors can be used in spectral regularization, where blanks in a space can be smoothed out by using the structure learnt from the rest of the space [17].

While the first type of generalization has been demonstrated, the second kind of generalization is more speculative. We propose that SR eigenvectors could theoretically support transfer where relationships between specific stimuli or states is preserved across tasks or contexts. This view has strong biological plausibility; instead of remapping in a new environment, like place cells, grid cell fields rotate in their orientation but keep the same relative relationship to each other [36]. Imagine you are in a new airport in a new country: although you have never been to this airport before, you know that they pretty much all have the same structure: check-in area, security, boarding gate. However, each given airport likely has a different orientation to how these basic components are laid out. A transferring of a grid structure from previous experiences of airports to tile the new airport layout can enable you to instantly transfer the associated actions and goals (e.g., scan your passport at a kiosk, take off your shoes at security, buy a snack at the shops next to the gate). This would predict that the grid code rotates along with the orientation of the new task context. So, if an item (e.g., the check-in counter) fell on the peaks of the grid pattern within a grid module in the first airport, then it would also do so in the second airport; similarly, if the security area fell at a trough in firing in one context, it would also do so in the second (Fig. 2B). A salient question with this view is whether an SR-based process in the hippocampus and entorhinal cortex could recognize whether different instances of kiosks at different airports, for example, belong to the same state. Such a process would be required to concern relations at a conceptual level (e.g., the idea of "kiosks" and "security areas") rather than at a literal physical level (specific kiosks and security areas). However, we think that this type of conceptual relationship could be plausible: as object categorization is already performed at the level of the inferior temporal cortex in the ventral visual stream [37], by the time the information arrives at entorhinal cortex and hippocampus, different instances of the same object type (such as kiosks) could already be recognized as a single state. Once in the hippocampus, representations have been shown to elide over details of specific episodes (such as distances) and instead demarcate more abstract ideas of "events" [38]. We further discuss in a later section ("How are states and transitions defined in SR?") how with a hierarchical process akin to feature selection driven by prefrontal processes, this generalization can become even more broad.

As a possible mechanism of how this could be implemented, a currently experienced successor map could be compared via a Chinese Restaurant Process to a belief distribution over successor map exemplars at every step [39], where if the current map matches one of the exemplars, then the exemplar map is updated; if it does not, it is categorized as a new exemplar map. This process can generalize better to novel maps and demonstrates biological plausibility by replicating several phenomena of place cells (e.g., "splitter cells" that fire in a location conditionally depending on previous history). With the addition of a transformation (such as rotation) to the eigenvectors of these successor map exemplars, relationships learnt from previous environments can be applied as a basis function for learning in the current one.

**SRs describe hippocampal and entorhinal function**

There has been growing evidence both that SR resembles the activity patterns of cells in the HC and EC in some tasks, as well as that findings in studies of animals and humans conform to the predictions of what you might expect from SR. For instance, cells that encode distance to goals in the HC of bats follow path distance, not Euclidean distance [40]. Another interesting study directly compared humans, rats, and RL agents that had to navigate to a goal within 25 different maze configurations, and where the starting position for each maze changed on every trial. The paper found that that SR (rather than planning using model-based tree search or model-free learning) best described the learning exhibited by the humans and rats[41].

Another recent study similarly sought to explain a variety of experimental findings with computational modelling of SR [42]. The authors modelled a variety of experimental paradigms, in which lesions to the HC or to the dorsolateral striatum (DLS) produced different patterns of behavior. For example, in a modified Morris Water Maze task, where the hidden platform was always accompanied by the same landmark that moved with it together, HC lesioned rats simply relied on the landmark to find the platform, indicating that they used a response-based strategy. In contrast, DLS-lesioned rats showed the opposite pattern, where they were impaired on the landmark-based strategy but showed improvement on learning the hidden platform over trials (indicating that they used an allocentric navigation strategy). The paper modeled the HC with an SR-based learning algorithm and DLS with a simple action-value learning strategy, and found that the SR could accurately replicate the consequences of lesioning the HC from animal data.

Another recent paper highlighted that SR is used in HC-EC representations in virtual task environments using fMRI in humans [43]. In this study, the participants performed random walks through a graph consisting of serially presented pictures (such as a key, an ear, a rabbit, or a motorcycle). The participants were not aware that these pictures had any underlying structure to them. The authors found that activity in the HC-EC system scaled with a measure of communicability closely related to the SR, which, importantly, was non-Euclidean – meaning that these areas represented frequency of visits that were experience-dependent.

## 3. Grid codes for structure abstraction of cognitive maps

**Grid codes as structural representations for cognitive maps and direct inferences**

Cognitive maps refer to an internal representation of an environment or a task structure consisting of relational information between entities or states [4]. One of the most powerful advantages of the idea of cognitive maps is that the animals can make a structural inference, draw a vector between states, and plan novel routes that they have not experienced yet. The grid codes may reflect these vectors for structure abstraction, that is, inferences of relationships between entities or states over an abstracted graphical representation of an environment or task's structural form. The abstracted graphical representation is built with nodes and edges specifying the transitions between elements (e.g. before and after in a sequence, in front or behind in navigation) or relationships (e.g. more than or less than on a number line; on top or below on a Lego tower), that is factorized from any specific contents.

During spatial navigation, the EC may provide a metric that enables animals to identify the coordinate of their current location and track self-motion during trajectories in an environment [8–12]. Specially, the spatial grids that describe how medial EC (mEC) grid cells tessellate 2D space are thought to facilitate path integration [44–49] for localizing one's position in the current environment. They also can be used for vector navigation [50–52] by computing direct vectors between locations in physical space. In principle, this mechanism can allow the animal to find an optimal route to reach the goal. Moreover, when an experienced path is blocked, animals have been shown to find an optimal alternative route to reach a goal even among those they have not explored, suggesting a cognitive map affords behavioral flexibility (Fig. 1D) [4]. Computational models have found that a deep neural network trained to perform path integration developed characteristic hexagonal grid codes, enabling agents to discover an unexperienced shortcut to the goal [24]. Moreover, under more biologically plausible constraints, a recent neural network trained to perform path integration showed that the robust hexagonal grid-like representations that they observed can be generalized outside of the training environment, for example to expanded environments [53]. These and other findings suggest that the grid codes may not be limited to encode experienced trajectories but also can be used to compute a direct vector between entities based on structural inferences about the relationships between locations in a 2D space [50,51].

Such a mechanism confers considerable computational advantages. In particular, by leveraging a rich structural representation of 2D topologies in the case of physical space, the grid code theoretically enables efficient computation of novel routes whose computational demands are not different from those for reinstating the previously learned relationships [1]. Therefore, planning with the direct vector to the goal in structural representations differs from the model-based planning using the tree search algorithm where the computational demand increases exponentially with planning steps [54,55], and also planning through precompiled routes and transitions using the SR [17]. Because the level of representation in SR depends on the amount of previous experiences, the behavior is limited to the pre-compiled or cached transitions between states. Therefore, it is challenging to account for how the HC place cells can reflect previously unexperienced routes during navigation [56] and reflect a shortcut to a goal that stitches together separately experienced routes [57]. Recent findings also suggest that the HC represents not only the successor states but distinguishes different states with identical sensory codes according to abstract regularities or hierarchical task structures, demonstrating that abstracted structural information is present in the HC [38,58–60]. Taken together, these findings suggest that the HC-EC system may use knowledge about the task or environmental structure to infer novel relationships. Explicitly representing the geometric structure of physical space or the transition structure of a graph could theoretically allow animals to infer relationships between any entities in the map and use direct vectors between entities to discover a novel route or solution.

If the HC-EC system builds a cognitive map not only based on a learning experience in one context, but also by integrating separate experiences learned from multiple contexts, the brain may construct a combined representation in a shared multidimensional space, assuming a common latent structure exists. Using the unitary cognitive map, the EC might employ the consistent grid codes to indicate the relationship between the same entities regardless of the current task demand. That is, according to this view the grid code's reference frame would be consistent for the same the latent structure of either a physical or task space, but would not be modulated by the current decision policy, choices, or selected actions. Instead, other brain areas, such as prefrontal cortex and (pre)motor cortical areas may receive inputs from the EC, translate this context-invariant structural code into the values for decision making or action selection in different contexts [61–64]. By providing consistent relational codes to other brain areas independently of the behavioral contexts, the grid representations may afford behavioral flexibility and facilitate the generalization across multiple decision-making problems. Notably, this idea is not incompatible with the recent findings showing the grid code adapts to the changes in environments. The grid codes are shifted in alignment to the goal location [29,30], and their firing fields are warped in response to the deformation of boundaries (e.g., a trapezoid environment). Importantly, these changes in the grid cell representation are not made immediately, in response to trial-by-trial changes, but progressively over time. This suggests that this change can be understood as a process of error correction of the internal representation of the latent structure or a process in adapting to fundamental changes in the environment or goals that impact the transitions sampled by the animal and therefore its internal representation [65].

**Factorized representations facilitate transfer learning across tasks**

When two tasks share the same structure or the common regularities of Euclidean space, recent theoretical proposals [3,6] predict that one can generalize their knowledge about one environment to make inferences about the other from sparse observations. To achieve this transfer learning, they proposed that the EC has a factorized representation of the cognitive map in which the abstracted representation of the structural code (i.e. scaffold, links, and edges; blue middle panel in Fig. 3D) is divorced from the sensory code (i.e. contents such as items or entities at nodes on the graph; red left panel in Fig. 3D). Specifically, the lateral EC (lEC), may represent the sensory code while the structural code may be represented in the medial EC (mEC), which contains grid cells. The theory proposing a factorized representation in the EC is based on the following points. First, the EC not only receives outputs from the HC, but also provides prominent inputs to the HC [66–69]. Second, when the brain perceives and processes the stream of sequential events from the outer world, the brain is thought to use two anatomically separate cortical networks that are preferentially involved in learning and representing the spatiotemporal context and items, respectively [70]. Specifically, the perceptual or semantic features are encoded in an anterior-temporal (AT) network that includes the inferior occipitotemporal cortex, perirhinal cortex, and lateral orbitofrontal cortex (lOFC) while the contextual associations that define relationships in a spatial, temporal, or abstract relational space are encoded in a posterior-medial (PM) network that includes retrosplenial cortex, posterior cingulate cortex, and medial PFC. Third, the PM and AT cortical networks have preferential connectivity to the medial and lateral subdivisions of the EC, respectively [70–73]. Thus, there are already functional distinctions between the inputs to lateral and medial EC that may be further factorized in EC. The computations and functions of the HC have been understood as binding entities or events into the contexts via the tri-synaptic pathway [74], which is consistent with this idea that the HC receives inputs from factorized representations in the EC and integrates them into a conjunctive representation. Together, these points suggest that the EC inputs provide the HC with highly structured representations of task environments. Because sensory inputs and latent states are learned separately in the factorized representation, the structural codes are easily separable from the sensory codes or from the original learning context.

Notably, the SR view proposes that grid codes emerge from the eigenvectors of the place cell population whose transition structure is learned directly from sensory transitions. Thus, factorization of structure and content would be difficult to achieve in the HC-EC system according to this view. This notion that grid cells emerge from place cell inputs into EC contrasts with the structural representation view that the input to HC from the mEC is already highly structured and fundamentally influences the place and other HC cell representations.

If the brain separately represents the structural codes from the sensory codes that are specific for each event, the representation of an entity in one context should be predicted from the activity patterns of other entities experienced in another context if two contexts share the same structural codes. In Morton et al. [75], participants learned the relationships between two sets of three stimuli using the same triad structure for associative inferences. They found that the activity patterns in the brain areas including anterior HC, parahippocampal cortex, medial prefrontal cortex, and lateral frontopolar cortex represent the relationships of each set with a similar geometrical representation, which subsequently allows predicting the pattern of an inferred relationship based on the expected location in the activity pattern space. Another study using the electrophysiological data acquired from non-human primates shows that the neuronal ensemble patterns of HC and the PFC neurons adopt a low dimensional geometry to represent each of the decision options efficiently according to the rule-dependent reward contingency [64]. These findings suggest that the brain constructs cognitive maps in a structure maximizing their generalizability. Specifically, the options in different contexts are represented with a similar geometrical topology in parallel planes if they are associated with the similar reward contingency (Fig. 3E). Therefore, the same linear classifier, such as a hyperplane, that was trained to classify options in one context can be applied to distinguish the options of a different context when paired with different or partly non-overlapping sensory stimuli.

The HC is thought to construct independent representations across environments while assigning the firing fields to specific places in a random manner referred to as 'global remapping' [76–78]. However, the proposed roles of the mEC representations as abstracting structure separated from the sensory contents provide an interesting prediction about HC global remapping. Whittington et al. [3] provided computational simulations suggesting that HC remapping is not random but predicted from the mEC grid cells' inputs, which maintain a consistent code across contexts that preserves the relationship between the same pairs. Indeed, they confirmed this prediction in simultaneous recording of place and grid cells across two different but structurally similar environments in each of two studies [35,79]. Furthermore, Chen et al. showed when an individual rat's place cell activity patterns from a T-maze navigation task were transformed into a common representational space using hyperalignment [80], the remapping in a given rat between right side trials and left side trials is predicted from the remapping matrix of other rats who also navigated the same maze [81]. This finding, therefore, suggests that the same spatial configuration of an environment is encoded similarly even in different rats, and it provided a generalizable input to the HC to produce a conjunctive representation. These studies further support the notion that while the HC has a distinctive neural representation for different environments or tasks, the EC may provide consistent grid codes if they share the same structure, allowing for knowledge to transfer from one task to another.

**Generalization for decision making in a novel task**

A factorized representation further suggests that the brain can use structural codes learned from different experiences and bind the current sensory inputs to the existing internal model. This enables one to generalize experiences in previous tasks into a novel task without needing to relearn the structure through experience. If the brain is aware of the relationships between entities in a novel task shares the same structural relationships (i.e. graphical representation of relationships) with a pre-learned task, it can apply the same decision policy to make structural inferences in the novel task (Bengio et al., 2012; McNamee & Wolpert, 2019; Sharpe et al., 2019) even when the sensory codes do not completely overlap with the previous one [82–84]. This transfer learning would not be explained if the representations of the task structure are limited to the SR since the sensory code is bound to the transition structure.

A recent study [85] showed that human participants discover a shortcut without direct experience, but instead from structural inferences. In this study, on day 1 in the 2 day experiment, human participants learned a relational structure by predicting the next sensory observation in a sequence generated from a probabilistic transition on a graph. On day 2, they performed the same task but with a different set of sensory stimuli, and crucially participants did not experience all possible transitions. In a following task, participants selected one of two stimuli which is closer to the target stimulus. While one of them is closer to the target in the structure, participants who do not make structural inferences would not know this since these stimuli had the same path lengths (number of transitions) to the target during day 2 experiences. This study showed that participants were able to make correct inferences, suggesting that participants found the unexperienced shortcuts between stimuli learned on day 2 by generalizing the structure learned on day 1, although it consisted of different sensory stimuli. Structure generalization is also observed in Zhou et al., [86] where the rats consecutively performed a series of odor sequence tasks that shared the same task structure, while a novel set of odor identities were introduced in each version of the task. They found that the rats learned the task structure faster each time they completed another version of the task. In addition to the behavioral findings, OFC neurons developed a low dimensional representation of the task structure faster when the rat could better generalize the task structure learned from previous task sets to the novel task. Recently, using fMRI, Baram et al. [87] showed that the EC activity patterns in human participants not only represent the correlation structure of a stimulus-outcome learning task, but also generalizes across different tasks using different sensory stimuli if they shared the same latent task structure. A recent finding using MEG also supports the representation of structure codes in medial temporal lobe factorized from sensory codes [88]. These findings show that the accurate representation of the abstracted task structure accelerates learning not only within a task but across tasks that preserve its structure. Taken together, these studies show that the brain reuses the structural codes selectively divorced from the sensory codes to afford behavioral flexibility and effective generalization, and that the EC plays a critical role in representing the structural code across task environments. Notably, the standard SR, which predicts the next sensory code based on the probability that one stimulus is followed by another, cannot readily account for these studies' findings showing generalization of the structural code dissociated from its learned sensory code.

**Grid codes for discrete decision making in non-spatial abstract task spaces**
Does the brain use the same grid codes to represent the relationship between entities in a non-spatial abstract space? In a physical space, the place coding typically stems from continuous multidimensional inputs, and the relationship between two places is learned from transitions over 2D spaces while receiving continuous sensory feedback (visual, vestibular). Therefore, the spatial relationships can be easily mapped onto multidimensional (2 or 3D) space. However, in many everyday decisions, the brain often learns the relationship between entities piecemeal in one dimension at a time (e.g., the number of rooms and the market price of a house, the levels of athleticism and the soccer IQ of an athlete, and the traits or ranks of people in a social hierarchy such as in the competence and popularity dimensions). Are abstract relationships sampled piecemeal reconstructed into a single multidimensional cognitive map even when it is not required for behavior? To examine the hexagonal grid code in an abstract space, we should first address what geometry the brain uses to represent a task structure when it was discretely sampled from separate experiences. The effects of vector angles in grid representations cannot be tested if the brain constructs and uses multiple 1-D maps instead of a unitary 2-D map. Park et al., [89] address this question: Does the brain construct multiple representations of the 1-D relationship of entities per learning context, or does it construct a cognitive map representing multidimensional relationships in a unified 2-D space? This study showed that the differences in activity patterns in the HC-EC system increases linearly as a function of 2-D Euclidean distances between entities in a combined space, even though the relationships in each of two dimensions are learned in different days separately through pairwise comparisons sampled out of sequence. The 2-D representation of the cognitive map in this study cannot be straightforwardly explained by the standard SR because the learning sequence was sampled out of order and the participants had never made any inferences that required combining the two dimensions (Fig. 4).

Given the finding showing that the HC-EC activity patterns reflect the Euclidian distances between entities in a 2-D abstract task space, Park et al. [22] used a similar training protocol and further tested for the hexadirectional grid-like representation characteristic of grid cell coding in the EC [20,90]. In this study, participants were asked to perform a novel decision task that is different from the tasks they had completed during training to learn the relationships between entities. This novel task requires participants to combine two 1-dimensional social ranks to compute the decision value on the fly, despite only learning each of two dimensions one at a time on different days during training. This study also found the grid-like representations for inferred direct vectors both in the EC and the interconnected mPFC and posterior cingulate cortex (PCC)/retrosplenial cortex (RSC), notably where grid cells have also been found in human single-cell recording during virtual spatial navigation [91]. Notably, the hexagonally symmetric patterns were still found in the EC and mPFC only for the trials in which the participants infer the relationships of the pairs that had never been compared before. This finding suggests the grid representations are generative, enabling bespoke computation of novel trajectories through abstract spaces on the fly. Whether these direct vectors are computed in mEC using a mechanism like path integration or vector navigation, or in PM regions that provide input to the mEC is unclear. However, hexagonal modulation for unlearned direct vectors shows that these novel inferences during on-the-fly decisions cannot be readily accounted for by the SR or its eigenvectors.

Another recent study [92] suggested that the grid-like codes in primate mPFC were implicated in computing decision values of novel options by integrating separately learned probability and reward magnitude dimensions. The macaques were trained to associate the probability and magnitude of rewards with the color and number of dots of the given visual cue (e.g., a darker blue color indicates high reward probability; the number of dots indicates a higher reward magnitude). During fMRI macaques made decisions between novel decision options that have not been shown during training. Their findings suggest that animals were able to infer the value of novel options correctly by building a cognitive map integrating information associated with the reward magnitude and the probability which were learned from separate experiences. Specifically, they found a modulation of hexadirectional grid-like codes in mPFC by the direction of the vector defined on a 2-D abstract value space which comprises two axes indicating the probability and the magnitude of expected rewards respectively. Additionally, when the mPFC was disrupted by transcranial focused ultrasound stimulation, they showed that the decision values were altered specifically when integration of the two dimensions was required to compute the value multiplicatively, suggesting an impairment of the ability to flexibly combine multidimensional information for decisions.

## 4. Limitations and potential solutions

**Limitations and advances in SR models to generalize beyond experiences**

We proposed that if grid cells indeed are the eigenvector of the SR, it would be able to generalize the learned relationships to some unexperienced transitions (if they are part of the same space), and novel positions of familiar stimuli with the same relationships. However, the grid code should still be subject to the limitations of the SR. As we discussed before, the SR is policy-dependent, and therefore it can only learn about transitions it has actually experienced. If a transition is impossible because of something like a barrier in the way, then there would be no SR representation and consequently no grid representation of that relationship.

One can distinguish between different scenarios where the SR is able to quickly adapt or not [93]. The SR adapts quickly to changes in value in an environment. Assuming a complete knowledge of the transition structure, the successor matrix is multiplied by the reward function of a state, so that if the location of a reward changes, the animal will not have to relearn where it is, but instantly be able to follow the route predicted in the matrix that allows it to go to the new location (Fig. 2C, middle). This allows for flexibility in learning in the face of changes. In contrast, the SR is not flexible to transition changes, such as if a route is no longer possible between two states. Therefore, if the transitions between states change, the SR will take longer to adapt (Fig. 2C, right).

Consequently, one difference between a SR-eigenvector-based model and other, structure abstraction models of the grid code is inference of novel routes and trajectories that have not been experienced, or possible. (As we allude to below, some advances to SR have allowed it to explore or infer off-policy trajectories, but a truly impossible transition – for instance, because of a blockade between two states -- will never be inferred, regardless of SR algorithm.) While an SR-based grid code can generalize learnt relationships or routes to some novel instances, it cannot infer the existence of relationships or paths that are not experienced. If, in a previously learnt route, a block happens (requiring a detour), the SR model would not be able take a new route if it has not been previously experienced or known. Other processes would need to kick in to figure out the new route, predicting longer reaction times during these "transition revaluation" trials. Suppose that in order to get from state 1 to state 3, one had to travel through a state 2 that is placed in an orthogonal direction to state 3. An SR would represent the trajectory as S1 → S2 → S3, and not as S1 → S3 directly (Fig. 3B), because a route directly between the two is not possible. In other words, in scenarios where the subgoals are in orthogonal directions to the ultimate goal, if grid-codes encode the eigenvectors of the SR, the grid representations should be aligned to the vector to subgoals but not to the direct vector to the ultimate goal [51].

However, the unembellished SR alone does not, by any means, need to be the complete model of the updating process in the HC-EC system. One issue with SR has been that if transitions are not completely known due to policy-dependent exploration, then a change in reward may still cause a large delay in updating. However, advances to the SR model have ameliorated this problem. Transitions that are not explored with a policy can be learnt or inferred with extensions to the SR model such as off-policy exploration during replay [94], a bank of possible successor maps that are matched to the present one based on reward function similarity [39], or control mechanisms that optimizes how the SR explores the transition matrix relative to a default policy [95]. In addition, as aforementioned, adding basis features analogous to border vector cells to SR learning can improve its generalization ability even further [33,96].

**How are states and transitions defined in SR?**

The SR, by itself, does not learn what the state space is itself – it simply learns over a state space it is given. Therefore, a key and seldomly addressed issue is how animals and humans learn and define the states themselves. This is one of the single most critical questions to how an SR-based learning system might be able to generalize to novel situations. If the state space is defined by simply the sum of the perceptual features of specific stimuli themselves, SR would not be able to generalize to novel stimuli that differed on these perceptual properties (although, as discussed above, object recognition processes in the visual stream as inputs to the MTL can likely already allow for different instances of the same object type to be categorized as the same state). However, if a state space is defined over the relevant features that allows for generalization, then the SR, by the simple virtue of those features, would be able to generalize more broadly. Successor features have been shown to achieve good generalization across changes in the transition functions, by compressing states that have the same reward prediction functions [97]. A hierarchical system whereby some process, whether located in the MTL or prefrontal cortex, could select relevant features and states, and then an SR-based system could learn over those features [98,99]. This hierarchical system is conceptually different than the factorized representation from the structural code perspective; instead of factorizing a relationship structure by the specific stimuli, this system would instead select a relevant feature (whether a simple one like a perceptual attribute, or a more abstract one like position), and SR would learn over those features.

A hierarchical feature selection process could plausibly help an SR learn state relationships. In Zhou et al., (2021), rats learn series of odors that make up two "arms" of a virtual maze. The two arms start off with distinct odors for the first two positions, but converge to the same set of odors thereafter. In the overlapping section of the odor sequence, two positions are differentially reinforced across the two arms (i.e., rewarded for positions 4 and not position 5 in one arm, and vice versa in the other arm). The rats learn this maze with the identity of the odors changed for each iteration of the maze. In that study, if we suppose that each state is a particular odor identity, then the SR cannot solve the problem of generalizing the sequence structure to other sets of odors because it represents the transitions between each odor. If the rodent has never experienced an odor sequence before, it cannot draw upon the policy-dependent SR to learn its actions. If instead, however, we suppose that the feature space selected was not odor identity but a feature based on sequence position and a selection rule based on the history of the first two odors experienced within the sequence, then SR could, in principle, transfer the transitions between the states in this task to a completely different set of odors (Fig.3C).

Plausible candidate brain regions for feature selection, and extracting more abstract features, are the orbitofrontal cortex (OFC) and medial prefrontal cortex (mPFC), together known as the ventral prefrontal cortex. Cellular recordings from Zhou et al. suggested that the OFC abstracted the task space [86], which is in line with other papers in humans that identifies this area as being important for representing a state space by directing attention to predictive features [89,100–105]. Moreover, recent findings provide evidences that the HC use the theta band (4~7 Hz) to transmit information about the changes in environments to the OFC and the stimulation suppressing the HC activity subsequently impaired the task structure encoding in OFC and flexible adaptation during goal directed decisions [106,107]. Neuroimaging work has identified that together with the HC, the mPFC abstracts task schemas in a slower process [108–115]. Although HC is a good candidate for fast mapping of associations between stimuli and initial state formation, elegant studies have demonstrated that the states it initially infers is limited to a single physical location (and not generalizable to multiple locations) [16,116]. The ventral frontal cortex, then, is likely needed for inference of a state space that is truly generalizable [1]. All of the previous mentioned studies involving conceptual spaces required extensive training [22,85,87,92,117,118], giving ample time for a state to be abstracted through a slower prefrontal process. Future studies should study the timing of abstract state inference and use tests of necessity to investigate the timing of when the OFC rather than HC is needed.

**Problem of learning and representing the structure of cognitive maps for structure abstraction from sparse experiences**

The SR can reduce the costs of online computation when the brain plans a route based on the transition matrix that is pre-compiled or cached during previous experiences. However, the standard SR won't afford behavioral flexibility when making decisions in a novel environment with new sensory stimuli even if it shares the same structure with a previously learned environment. Because the transition structure is learned from the probability that one sensory observation (or state) is followed by another in the SR, the brain cannot have a separate graph structure divorced from its sensory codes, which limits the generalization of structural knowledge across different tasks. Contrary to the SR, knowing the underlying structure of abstract tasks or relationships would allow the brain to make accurate inferences before direct experiences. Moreover, when two independently learned types of knowledge share the same graph structure, the brain can take advantage of same policy to make decisions or inferences in a novel task that has not yet been experienced. However, while the SR provides a mechanistic understanding of how the brain can learn a global structure from sequential sensory observations, it has been less clear about how the brain can build a global representation of cognitive maps from sparse and limited experiences. Compared to the relatively cheap costs of online computation to build and generalize with a SR, building and planning over an on-line representation of a global task structure requires considerably more computational resources.

To circumvent heavy computational cost required for online learning, the brain might adopt two solutions. First, the brain might have a set of schemas, or basis forms of structural knowledge, that could be learned from the previous experiences and find the one that is analogous or similar to the observed transitions to scaffold it to the new sensory environment. By providing a prior for Bayesian inference, the pre-built structural knowledge allows the brain to make structural inferences based on the current belief of what would be the global task structure. Kemp et al. have proposed a generative Bayesian model in which a wide variety of graphs, including complex forms, can be generated from a combination of basic graph forms [119,120]. Mark et al. [85] built a simulation to show that a minimal number of basis set structures can be employed and developed to represent a series of sophisticated structures. Furthermore, they provided evidence showing that the participants who may use a less effective structure as the prior took longer to learn a global structure, illustrating the role of the prior on structural forms in shaping predictions. Moreover, Plitt et al.,[121] demonstrated that HC remapping patterns can be differentiated by the level of certainty that an animal has experienced in previous blocks, suggesting that building a cognitive map and remapping may be driven by structural representation inputs, which reflect the current belief about the task structure.

Second, to distribute the burden of online learning, the brain might learn task structure offline. Computational models [122,123] have proposed that a physiological phenomenon known as 'replay', which refers to compressed patterns of activity reflecting both previously experienced or to-be-taken sequences of states, might be a key mechanism for learning a given task structure. Previous studies have shown that the memory consolidation process during sleep is critical for learning structural relationships [124]. In addition, Stella et al. [125] showed that the replay sequences during sleep do not follow the order of experiences but were sampled in a random fashion. Recently, Liu et al. provide evidence that 'replay' may play a critical role for transferring structural knowledge learned from one task to the other [88]. In this study, human participants were asked to reconstruct a series of sensory stimuli according to a certain order that they had learned in a different sensory environment. In their task the identity of items was learned independently from the correct sequence order, meaning that the sensory code that were relevant to the current task were dissociated from the structural code. Using magnetoencephalography (MEG), they decoded the replay of non-spatial sequences that the brain made during a rest period between tasks. Specifically, they showed that the brain transfers structural codes to reconstruct the replay sequence of the task-relevant sensory codes. Moreover, the decoding results showed that the position in the structural information preceded the content of the sensory code, supporting the idea that the factorized representation affords generalization. Another study [126] showed that the brain replays the sequence associated with their current choice, but also replays the sequences that are not shown in the current trial but share the same latent cause to lead to the identical current outcome in the reverse order. That is, the brain uses a cognitive map to choose the sequence to assign credit to, which is inferred from recent updates to the reward contingency, even when these sequences are only experienced remotely from the current trial. These findings suggest that the brain employs offline replay to build, update, and transfer structural knowledge. The idea of distributing computational costs with offline replay is similar to SR-DYNA. However, while SR-DYNA assumed that replay sequences are sampled randomly, recent findings suggest that the brain can use structural knowledge to guide which replay sequence should be prioritized. Therefore, the replay sequence is not sampled by temporal adjacency, but rather the causal relationship defined by the internal model, which could even be learned from separate experiences. These findings suggest that replay might play a vital role in the brain to build a task structure beyond the SR which we predict may guide the EC to discover previously unlearned relationships.

## 5. Future directions

In considering the utility of the grid code in generalization, it is important for future work to define models of grid cells with clear, falsifiable predictions. Both SR and structural abstraction perspectives can have adaptations and added mechanisms that allow them to perform inference beyond the respective bare-bones versions, but we believe it is important for these theoretical perspectives to guide experimentalists' hypotheses about where these additional mechanisms may be implemented in the brain, and what the behavioral consequences of those might be (e.g., higher RT, or a requisite break period, for mechanisms requiring replay). In particular, it is worth considering some of the following points when designing studies intended to model the grid code.

**When is a grid code not useful?**
So far, there has been limited discussion of the limitations of a grid code in generalization and conceptual navigation. This is largely due to the fact that the conceptual tasks investigating the grid code are carefully designed to elicit them. Task demand may change the representations in the same brain regions where a grid code has been found, even given the same task space. One instance in which the grid code may not appear, even given a 2D stimulus space, may be situations where a different sort of representation may be more efficient, as in a one-dimensional subjective value signal. Lee, Yu et al. [127] applied the same grid code fMRI analysis used in previous physical and conceptual navigation studies [20,90] to the data in a canonical delay discounting task. The task was not designed to study the grid code but share the same two-dimensional characteristic of a stimuli space as other conceptual navigation studies; in fact, a theoretical grid code in the stimulus space in that study is highly correlated with the subjective value signal. However, the study did not find any evidence of a grid code in that study. Instead, a subjective value signal was the better explanation for the neural code in vmPFC. A similar study in macaques [92] found a grid-like code in mPFC when the value-based stimuli are presented sequentially, but not when a choice was required between the options presented together on the screen. These studies together suggest that the grid code may not inherently be necessary for value-based decisions but could instead be task or process dependent – if a subjective value representation is sufficient (as in Lee et al.) a grid code may not be used during value-based choice (even if a grid code is used to represent the 2D stimulus space itself). Future studies can carefully tease apart scenarios in which a two-dimensional grid code may or may not be useful for the task requirements.

**When and how do grid representations become useful?**
Many everyday decisions require one to integrate multi-dimensional information into subjective values as a comparable common currency. When one needs to evaluate the value of a novel decision option, recent findings show that the brain could use the grid code to determine the relative position of the option in an abstract attribute space [22,92]. That is, the brain not only stores the value of experienced decision options but can also construct a cognitive map that best captures the regularities underlying an abstract task and utilize the cognitive map generatively to construct the value. During training in both Bongioanni et al. and Park et al., participants only learned the attributes of decision options in one dimension at a time. Therefore, subjects could have adopted multiple 1-D representations of attribute values instead to represent learned relationships in each dimension separately (e.g. for popularity and competence attributes or probability and magnitude attributes). However, these findings suggest that the brain builds a unitary representation in a multidimensional space by integrating information learned from different experiences, even when this is not necessary to solve problems during initial learning or training. Importantly, Park et al. found the brain constructed a 2-D representation even when subjects did not know they would have to subsequently use this representation for generalization to make novel inferences.

Moreover, these studies found that the brain has grid-like representations of the *direct vectors* between novel decision options in the abstract 2-D attribute space (Fig.4). By having grid representation over the 2-D cognitive map, first, the brain could construct the location of the novel decision options and relate the novel option to previously learned relationships, thus allowing one to generalize previous experiences to guide future decisions [92]. Second, the brain can translate the relative positions between entities in 2-D attribute space into a subjective value for a novel decision-making task even when it differs from the decision values computed for the training task [22]. In addition, Bongioanni et al. showed that transcranial ultrasound stimulation of mPFC that potentially interrupted grid representations and altered subjective values of novel decision options and subsequent choices when they required integration of both dimensions. Likewise, Park et al. showed that the activity of brain areas including mPFC encoding trial-by-trial decision values also showed grid-like representations in alignment to the EC grid orientation, suggesting that the EC grid representation may provide an input to value computation when it is based on a cognitive map. Taken together, these findings suggest that the brain constructs a task representation that can be generalized across multiple contexts, and that grid representations are useful when integrating multidimensional information in the abstract task space, particularly when computing values of novel decision options that have not been experienced before. We hypothesize the structural abstraction reflected in the grid representation makes possible both efficient localization in task spaces [44], using path integration [46,47], and for the online construction of direct inferences. Whether these direct inferences are computed locally in the EC or in concert with interconnected mPFC and OFC remains an important open question. Furthermore, whether the brain uses the same grid code to identify the relationship between identical items or states in different behavioral or decision making contexts (e.g. When inferring the relative social hierarchy between two people in the competence dimension and inferring that in the popularity dimensions, whether the brain uses the consistent grid codes to indicate the relationship between two people in the abstract social space?), and how the brain may translate the context-invariant/ task-agonistic grid code into different decision values according to task demands in the current context remains unclear (e.g. When one is higher in the social hierarchy in the competence dimension than the other but lower in the popularity dimension, how do the same grid codes produce different outputs of decision making or action selection?).

**Complex environments**
In order to tease apart the contributions of the SR vs. other mechanisms that could produce a grid code, it is necessary to use more complex environments that are large and contain multiple barriers. The SR, as we have mentioned, would best represent path distance rather than Euclidean distance. Smaller and open spaces are best represented by a Euclidean cognitive map that is reflected in the grid code, but this can be consistent with both SR and vector navigation views. In contrast, cognitive graphs are used more often for cluttered and large spaces [128,129]. Boundaries cause distortions in the grid code (e.g., Carpenter et al. [130]) and therefore it is important for future experiments to test the contributions of each model best with boundaries. Specifically, do participants represent space or conceptual relationships in a way that is more SR-like vs. Euclidean-like in a space with more boundaries, or when their trajectories are forced in a certain direction? And how would that affect their generalization to novel instances?

It is also worth considering the fact that, unlike a computational agent, biological organisms like rodents and humans can gain knowledge of spatial trajectories that they have not traveled without any inference process required by simply using their senses (i.e., if I see a path to my right, I know it's a possible route even without having gone down it; if I'm in an open field, then I know all directions are possible even if I have not physically traveled in every direction). Thus, future studies that seek to test Euclidean vs. path representations in the brain should take into account the spatial trajectories obtained via visual (or other sensory) knowledge, for example by "blinding" their participants to any other possible path but the one they are on.

**Grid-codes on high-dimensional cognitive maps**

How does the brain balance the benefits of high dimensional representations with the benefits of generalization and behavioral flexibility of the low-dimensional grid code representation? On the one hand, if the brain constructs a low dimensional representation while incorporating only the variables relevant to the current task demands, the representation will no longer represent factors that could become important in the future. Moreover, previous findings suggest that the brain not only learns the values of decision options in the current task but also allocates substantial resources to learn the regularities across tasks and to represent a global task structure that can be generalized across multiple contexts of decision making [60,86,89,102,131]. On the other hand, a single high dimensional cognitive map increases the needs for cognitive control to select information relevant to the current task while deprioritizing other information in task irrelevant dimensions. Collins et al. suggests that the brain may store multiple sets of rule representations, and the prefrontal cortex (PFC) represents one of them at a time [98,132,133]. The brain may infer the representation that best explains the current observations while benefiting from generalization within the set. In doing so, the neural representation of multiple sets of rules could provide efficient value computations for decision making in a specific task. A recent theoretical model [26] proposes that if the high dimensional representation constructed in the HC provides a low dimensional projection to the EC. Subsequently, the grid codes could be employed for making inferences of relationships of entities on the low dimensional manifold that represents the relationships between entities according to their values in task-relevant dimensions. Note that the high dimensionality representing multidimensional variables in a complex task [134] should be dissociated from noisy representations which also have high dimensionality [135,136]. By extension, other cortical areas such as mPFC and OFC may also receive inputs from the HC-EC system and represent a structure with reduced dimensionality while the HC continuously updates a high dimensional representation. In this way, the PFC, for example, might also have a structural representation facilitating efficient computation of task-relevant variables [62–64]. If true, then the representational structure of different brain areas will help to understand what variables are computed and how they are engaged with different cognitive functions. Future studies of the computational mechanisms deciding the dimensionality, graphical representations, and levels of abstraction in the cognitive maps would be important to understand how the grid code affords both generalization and flexible decision making.

## Acknowledgments

The authors MRN and LQY would like to thank Drs. Avinash Vaidya and Ida Momennejad for productive conversations on ideas in this paper. The authors also gratefully acknowledge support from National Institute of Health award R00AG054732 to MRN and a subaward from the NIH grant 1P20 GM103645 to MRN and LQY. This work was also supported by an NSF CAREER Award 1846578 and an R56 MH119116 from NIH awarded to EDB.

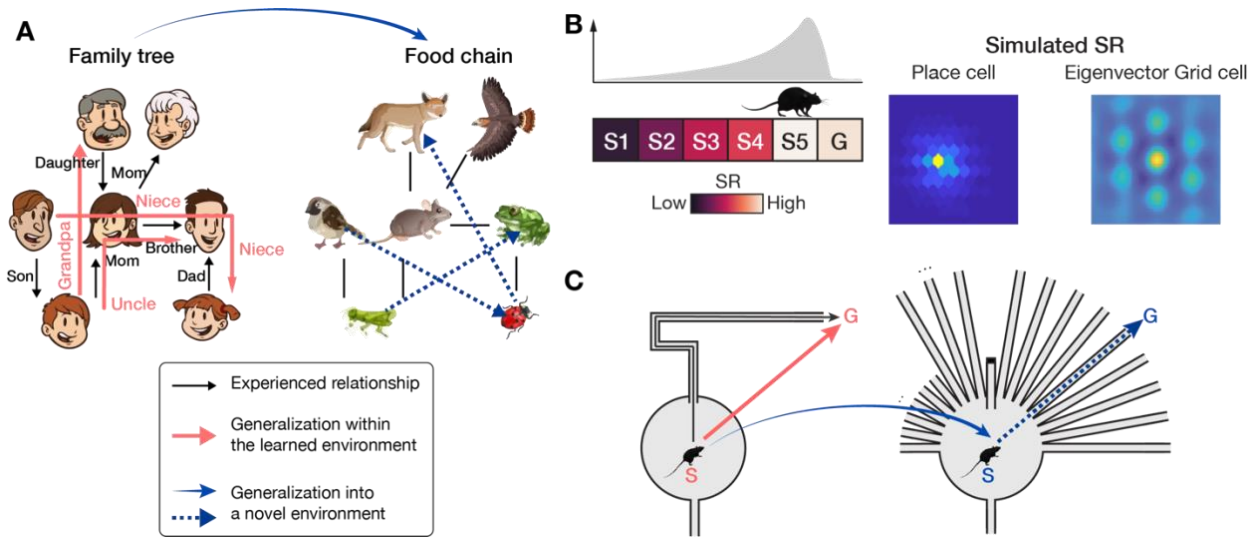

**Figure 1. A.** The brain organizes not only spatial but also non-spatial relationships in a structural form and construct cognitive maps. Theoretically, cognitive maps allow the brain to generalize learned relationships (black arrows) to make inferences of unlearned relationships (red arrows) defined with the dimensions of the same task space (within-task generalization). Moreover, when two cognitive maps share the same graph representation, the brain can generalize the policies or the methods to make inferences about another novel task without needing feedback (e.g., family tree network to food chain network; blue arrows for the between-task generalization). **B.** A successor representation (SR) along a linear track, representing discounted expected occupancy over states (left). This representation can reproduce the firing pattern of place cells and, by taking the eigenvector, grid cells (right). **C.** Tolman et al.[4] showed that after having a reward, a rat not only reinforces their actions that led to the reward (black arrows) but also learned the spatial relationship between starting state (S) and the goal (G). When the rat is introduced to a new environment (right) where the previous route leading to the goal is blocked, rats use the learned relationship from the previous environment (blue arrow) and are more likely to choose the optimal route to the goal (blue dotted arrow) without experiences in the novel maze, supporting the idea that structural knowledge is critical for generalization across different environments.

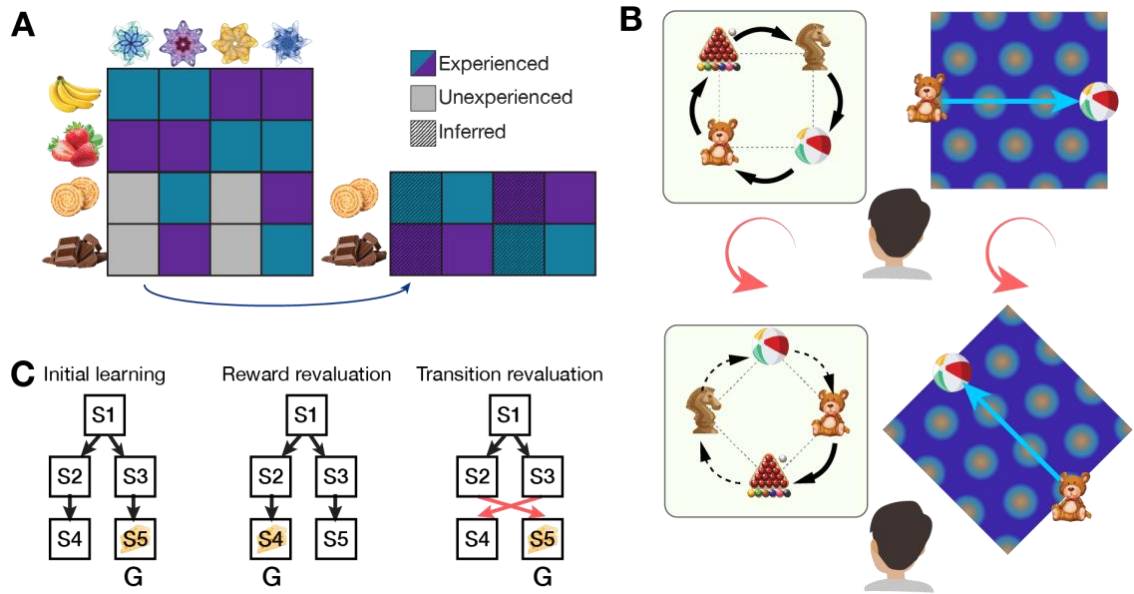

**Figure 2. A.** Generalization to unexperienced transitions via spectral regularization, using SR eigenvectors. In this transition matrix, the column states transition to the row states. The colored cells are the experienced transitions (blue = transition possible; purple = no transition possible), and the grey ones are the ones that must be inferred. The eigenvectors of SR fill out the rest of the matrix: in a situation where transitions have been experienced and can be summarized by a set of eigenvectors in one subset of stimuli (top half of matrix), and in which some of the same transitions have been experienced between the other subset of stimuli, the eigenvectors can be used to fill out the rest of the matrix without having been experienced (grey cells). **B.** Generalization by transferring structure to a rotated version of the same states. The EC grid code could theoretically rotate with the task space, transferring relationships between the stimuli, even if the absolute locations have changed (i.e., if the man knows that the teddy bear transitions to the beach ball, then the grid code would be able to maintain this relationship on the same activation peaks, whether the transition requires navigating to the east or to the northwest). **C.** An example of different types of revaluation given changes in the environment. After learning the initial transition structure (left), with the reward (cheese) located in S5, an SR would be able to instantly navigate to the new reward location if it has changed (middle). However, SR would not be able to quickly learn if the transition structure is changed (right).

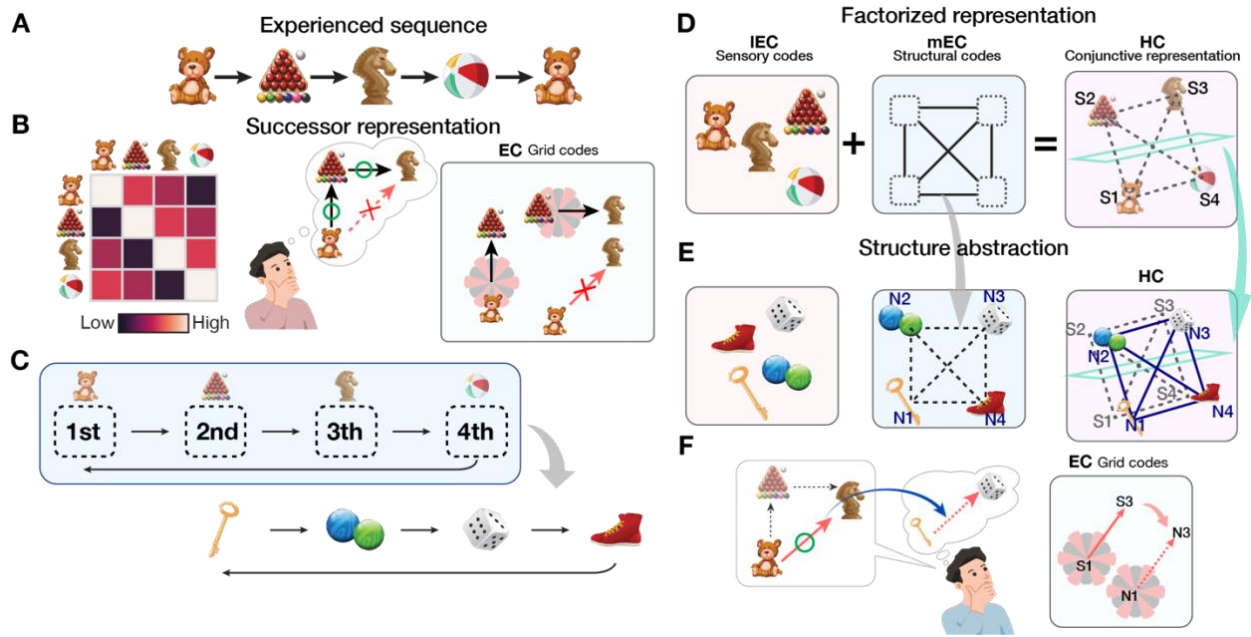

**Figure 3. A.** An example of an experienced sequence of stimuli. **B.** An SR matrix representing the experienced transitions between the stimuli (left). The limitation is that an SR would not be able represent an unexperienced trajectory. Therefore, the diagonal trajectory between the teddy bear and the chess piece cannot be inferred by the SR or a grid code based on it. **C.** Proposed mechanism for which SR can be learnt over features. If the states are defined over the relevant features (positions in sequence) rather than by the particular stimuli (teddy bear, pool ball set, chess piece, beach ball), then the transitions can be applied to novel sets of stimuli (key, marbles, dice, shoe). **D.** The factorized representation. While experiencing a sequence of events (Fig.3 A), the brain not only can learn the objects encountered (sensory code; left) but also their relative positions (structural code; middle) while integrating paths to infer the global structure. The EC has been proposed to encode the structural and sensory codes separately in the medial and lateral EC respectively [3,6]. The HC binds the structural and sensory codes resulting in a conjunctive representation (right), which can reflect the geometrical relationships of the task structure. **E.** The factorized representation, in which structural codes are not bound to a learned context, allow the brain to use a structure learned from different tasks to learn and infer the task structure of a novel task. Right panel: This high dimensional geometry representing the task structure allows the same linear classifier trained from the previous task to also be applied to a novel task (e.g. the green hyperplane classifies not only S1 and S4 *vs.* S2 and S3 but also N1 and N4 *vs.* N2 and N3). **F.** The brain may afford structural inferences by applying similar grid codes over the pre-learned structure to related entities encoded by novel sensory codes in different contexts. Right panel: the grid codes indicating S1→S3 could guide inferences of an unlearned relationship, N1→N3).

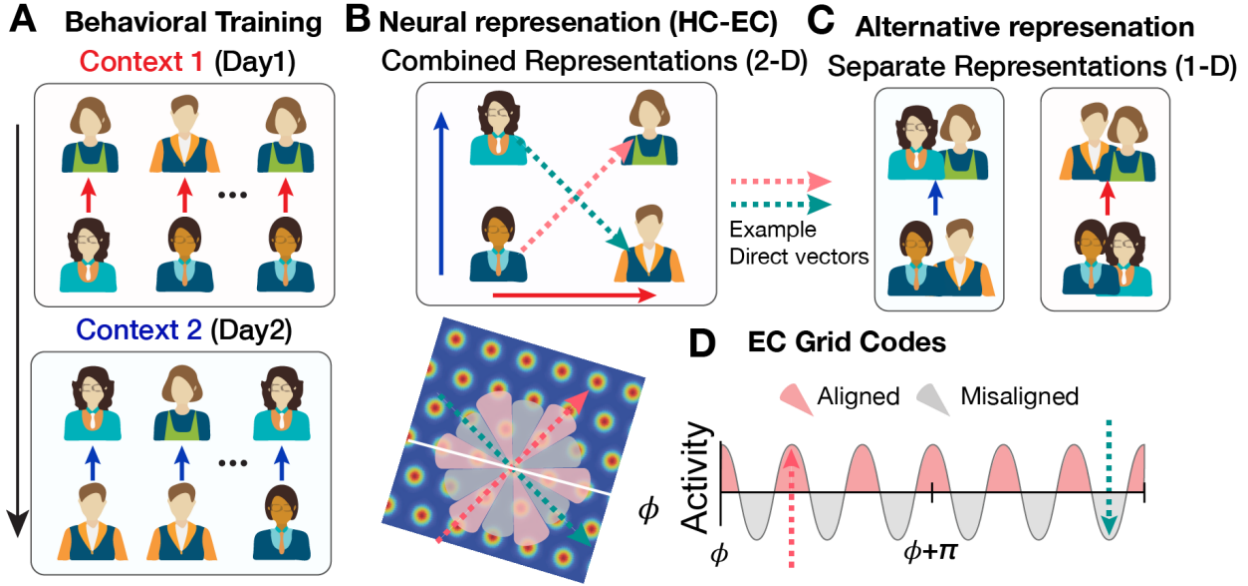

**Figure 4. A-C.** After learning social hierarchy relationships between discretely sampled individuals in one of two dimensions (popularity and competence) in different days (**A**), Park et al., showed that the HC, EC, and OFC integrate the piecemeal learned relationships into a combined 2-D structural representation[89] (**B**) even though the brain could have alternatively built two 1-D representations per dimensions (**C**). **D.** When participants needed to compute decision values for a novel task (different from the task used for initial learning) based on the relative social hierarchy ranks between individuals in *both* dimensions, activity in the EC (and medial prefrontal cortex, posterior cingulate cortex, and temporoparietal junction area, among other areas) was modulated by the angle of the inferred direct vector between individuals on 2D social space in six-fold symmetry [22].